 \documentstyle[11pt]{article}

\begin{document}

 \title{Comment on ``Noncommutative gauge theories and Lorentz symmetry,
Phys. Rev. D {\bf 70}, 125004 (2004) by R.Banerjee, B.Chakraborty and K.Kumar,
''}
 \author{Alfredo Iorio \thanks{E-mail: iorio@ipnp.troja.mff.cuni.cz} \\
 Institute of Particle and Nuclear Physics, Charles
University of Prague  \\ V Holesovickach 2, 182 00 Prague 8 -
Czech Republic \\ and \\
Department of Physics ``E.R.Caianiello'' University of Salerno and INFN \\
Via Allende 84081 Baronissi (SA) - Italy}
\date{\today}
\maketitle

\begin{abstract}
We show that Lorentz symmetry is generally absent for noncommutative (abelian) gauge theories and obtain a
compact formula for the divergence of the Noether currents that allows a throughout study of this instance
of symmetry violation. We use that formula to explain why the results of ``Noncommutative gauge theories
and Lorentz symmetry'', Phys. Rev. D {\bf 70}, 125004 (2004) by R.Banerjee, B.Chakraborty and K.Kumar,
interpreted there as new criteria for Lorentz {\it invariance}, are in fact just a particular case of the general
expression for Lorentz {\it violation} obtained here. Finally, it is suggested that the divergence-formula should
hold in a vast class of cases, such as, for instance, the Standard Model Extension.
\end{abstract}

\bigskip


\bigskip

\setcounter{equation}{0}

We want to illustrate here why the conclusions of Ref.
\cite{Banerjee:2004ev} on the possibility to preserve full
Poincar\'e invariance for abelian noncommutative gauge theories
(NCGTs) {\it \`a la} Seiberg-Witten \cite{Seiberg:1999vs}, described by a
lagrangian of the form
\begin{eqnarray}
\hat{\cal L} = - \frac{1}{4} \, \hat{F}^2  =
\frac{1}{4} F^2 + \frac{1}{8} \theta \cdot F \, F^2
- \frac{1}{2} (F \theta F) \cdot F + \cdots \equiv \hat{\cal
L}|_{O(\theta)} + \cdots \;, \label{firstorder}
\end{eqnarray}
are incorrect. Our notation is standard: $\hat{F}^2 = \hat{F} \cdot \hat{F} = \hat{F}_{\mu
\nu} \hat{F}^{\mu \nu}$, and so on, where $\hat{F}_{\mu \nu} =
\partial_\mu \hat{A}_\nu - \partial_\nu \hat{A}_\mu - i (\hat{A}_\mu  \star \hat{A}_\nu -
\hat{A}_\nu  \star \hat{A}_\mu)$ ($F_{\mu \nu} =
\partial_\mu A_\nu - \partial_\nu A_\mu$) is the noncommutative
(commutative) field strength, and $\theta_{\mu \nu} = - i ( x_\mu
\star x_\nu - x_\nu \star x_\mu )$, is the $x$-independent
antisymmetric matrix encoding noncommutativity of coordinates.
Namely, based on the results of Ref. \cite{Iorio:2001qy}, we shall
explicitly show that what in Ref. \cite{Banerjee:2004ev} are
interpreted as novel criteria for Lorentz invariance - e.g.,
$\partial_\mu M^{\mu \nu \lambda} = 2 [(\delta \hat{\cal L} /
\delta \theta_{\alpha \nu}) \theta^\lambda_{\, \alpha} - (\nu
\leftrightarrow \lambda)]$, cf. Eqs.(81) and (82) in
\cite{Banerjee:2004ev}  - are in fact the opposite.

In Noether's first theorem \cite{emmy},\cite{Brading:2000hc} the action ${\cal
A} = \int d^4 x {\cal L} (\Phi_i, \partial \Phi_i)$ is said to be \textit{invariant}
under the infinitesimal continuous transformation $\delta_\epsilon$ (or, equivalently,
$\delta_\epsilon$ is said to be a \textit{symmetry} of ${\cal A}$) - here
$\{ \Phi_i (x)\}$ is the set of fields of any spin-type, $i$ is a multi-index
and, although the theorem holds for the general case, for the case
in point we need only to consider first derivatives of the fields
- when, for all field configurations (off-shell), $\delta_\epsilon {\cal A} = 0$.
If this happens then there is a conservation law
\begin{equation}\label{1}
    \partial_\mu J_\epsilon^\mu = \sum_{\Phi_i} \Psi [\Phi_i] \delta_\epsilon \Phi_i \;,
\end{equation}
when the field configurations respect $\Psi [\Phi_i] = 0$ (on-shell). Here $J_\epsilon^\mu$ is the current
for a rigid gauge transformation, $J_\epsilon^\mu = \sum_{\Phi_i} \Pi^{\mu \, i} \delta_\epsilon \Phi_i$, or
for a spatiotemporal transformation (including supersymmetry),
$J_\epsilon^\mu = \sum_{\Phi_i} \Pi^{\mu \, i} \delta_\epsilon \Phi_i - {\cal L} \delta_\epsilon x^\mu
( + V^\mu)$, $\Pi^{\mu \, i} = \delta {\cal L} / \delta \partial_\mu \Phi_i$  and
$\Psi [\Phi_i] = \partial_\mu \Pi^{\mu \, i} - \delta {\cal L}/ \delta \Phi_i$  are, in Noether's terminology, the
``Lagrange expressions''.
There are further possibilities for conservation in certain special cases, i.e.
when although for some $\Phi_i $ $\Psi [\Phi_i]$ is not zero the corresponding
$\delta_\epsilon \Phi_i$s on the right side of (\ref{1}) can be set to zero without
this producing a vanishing current $J_\epsilon^\mu$ on the left side. This happens for special
choices of the parameters and only for certain theories, like e.g. the
theory (\ref{firstorder}) in point. In what follows we shall call these ``relic symmetries''.

In any case, invariance of a classical field theory under the continuous
transformation $\delta_\epsilon$ always means
\begin{equation}\label{2}
    \partial_\mu J_\epsilon^\mu = 0 \;.
\end{equation}
Simply on the basis of this, those results of Ref.
\cite{Banerjee:2004ev} that say $\partial_\mu J_{\rm Lorentz}^\mu \neq 0$,
can never be interpreted as an invariance.

Let us now consider the infinitesimal Poincar\'e transformations as $\delta
x_\mu = - f_\mu$, with $f_\mu = a_\mu$ and $f_\mu = \omega_{\mu \nu}
x^\nu$ for infinitesimal translations and homogeneous Lorentz
transformation, respectively. According to their indices
structure, the fields $\{ \Phi_i (x) \}$ respond to those
coordinates changes as $\delta_f \Phi_i = \Phi_i (x) - \Phi'_i (x) = {\bf L}_f \Phi_i (x)$ (see, e.g.,
\cite{jackiw}).
Note that the changes are evaluated at the same point $x$. This is not strictly necessary
but simplifies the analysis because $[ \partial_\mu , \delta_f ] =
0$. Here the Lie derivative along the vector $f^\mu$ has the
usual expression
\begin{equation}\label{lie}
{\bf L}_f \Phi_{\mu \dots \nu}^{\lambda \dots \kappa} = f^\alpha
\partial_\alpha \Phi_{\mu \dots \nu}^{\lambda \dots \kappa} +
\Phi_{\alpha \dots \nu}^{\lambda \dots \kappa} \partial_\mu
f^\alpha + \cdots + \Phi_{\mu \dots \alpha}^{\lambda \dots \kappa}
\partial_\nu f^\alpha - \Phi_{\mu \dots \nu}^{\alpha \dots \kappa}
\partial_\alpha f^\lambda - \cdots - \Phi_{\mu \dots \nu}^{\lambda \dots \alpha}
\partial_\alpha f^\kappa \;.
\end{equation}

The ten currents can be written in the compact form
\begin{equation}\label{jpoinc}
    J_f^\mu = \sum_{\Phi_i} \Pi^{\mu i} \delta_f \Phi_i - {\cal L} f^\mu \;,
\end{equation}
where $J_f^\mu = T^{\mu \nu} a_\nu$ and $J_f^\mu = M^{\mu \nu
\lambda} \omega_{\nu \lambda}$ for translations and Lorentz
transformations, respectively, with $T^{\mu \nu}$ the canonical energy-momentum tensor
and $M^{\mu \nu \lambda}$ the angular momentum tensor. We can call $\delta_f \Phi_i$ the
``algebraic'' transformations, i.e. the transformations purely
based on the indices structure of the fields, as opposed to
those generated by the Noether charges $\Delta_f \Phi_i =
\{ \Phi_i , Q_f \}_{\rm Poisson}$, where $Q_f = \int d^3 x
J_f^0$, which we call ``dynamical'' transformations. For consistency,
the latter can only coincide with the algebraic transformations or be zero
\cite{Iorio:2001qy}: $\Delta_f \Phi_i = \delta_f \Phi_i$ or
$\Delta_f \Phi_i = 0$.

Suppose now that there are only two fields, $\{ \Phi_i \} =
(\phi_j, \chi_k)$, and that the field $\phi_j$ is dynamical, i.e.
the relative $\Pi^{\mu \, j}$ is nonzero, while the field $\chi_k$
is non-dynamical, i.e. $\Pi^{\mu \, k} = 0$ (as before, $j$ and
$k$ are to be understood as multi-indices). The currents do
not contain the algebraic variations $\delta_f \chi_k$
\begin{equation}\label{4}
J_f^\mu = \Pi^{\mu j} \delta_f \phi_j - {\cal L} f^\mu \;,
\end{equation}
thus they cannot depend on whether the field $\chi_k$ has been
varied in the action. We want to study now the invariance and
dynamical consistency properties of theories of this class. We shall
do that by studying $\partial_\mu J_f^\mu$.

In general, we cannot say how $\partial_\mu J_f^\mu$ looks like.
This depends on the way $\chi_k$ appears in the action. For
instance, it could be fully decoupled from the dynamical field
$\phi_j$, in which case no sign of it would be found in $J_f^\mu$,
hence in $\partial_\mu J_f^\mu$ (see, e.g., \cite{Iorio:1999yx}).
Let us consider instead the lagrangian for two vector fields
$\phi_j = B_\mu $ and $\chi_k = P_\mu $
\begin{equation}\label{B}
   {\cal L} = \frac{1}{2} \partial_\mu B^\alpha \partial^\mu B_\alpha - V(B)
    + B_\alpha P^\alpha  \;,
\end{equation}
where $V(B) = a B^2 + b B^4 + \cdots$ and the non-dynamical field
is indeed coupled to the dynamical one to form what {\it would be
a scalar} if both fields are transformed according to their
indices structure (algebraically). We have $\Pi_B^{\mu \nu} \equiv
\Pi^{\mu \nu} =
\partial^\mu B^\nu$, $\Pi_P^{\mu \nu} = 0$, $\Psi [B_\nu] =
\partial_\mu \Pi^{\mu \nu} + \delta V / \delta B_\nu - P^\nu$,
$\Psi [P^\nu] = B_\nu$, $\delta_f B_\nu = f^\alpha
\partial_\alpha B_\nu + B_\alpha \partial_\nu f^\alpha$ and
$\delta_f P^\nu = f^\alpha
\partial_\alpha P^\nu - P^\alpha \partial_\alpha f^\nu$. The Poincar\'e
currents are $J_f^\mu = \Pi^{\mu \nu} \delta_f B_\nu - {\cal L}
f^\mu$ and (using $\partial_\mu f^\mu = 0$ and $\partial^2 f^\mu =
0$)
\begin{eqnarray}
\partial_\mu J_f^\mu & = & (\partial_\mu \Pi^{\mu \nu}) \delta_f
B_\nu +  \Pi^{\mu \nu} \partial_\mu \delta_f B_\nu - f^\mu
\partial_\mu {\cal L} \nonumber \\
& = & (P^\nu - \delta V / \delta B_\nu) ( f^\alpha
\partial_\alpha B_\nu + B_\alpha \partial_\nu f^\alpha ) \nonumber \\
& + & \Pi^{\mu \nu} [(\partial_\mu f^\alpha) \partial_\alpha B_\nu
+ f^\alpha \partial_\mu \partial_\alpha B_\nu + (\partial_\mu
B_\alpha ) \partial_\nu f^\alpha] \nonumber \\
& - & f^\mu [(P^\alpha - \delta V / \delta B_\alpha) \partial_\mu
B_\alpha + \Pi^{\alpha \beta} \partial_\mu \partial_\alpha B_\beta
+ B_\alpha \partial_\mu P^\alpha] \nonumber \\
& = & B_\alpha (- f^\mu \partial_\mu P^\alpha + P^\nu \partial_\nu f^\alpha ) \label{secondlast} \\
& + & (\partial^\mu B^\nu) (\partial^\alpha B_\nu) \partial_\mu
f_\alpha + (\partial^\mu B^\nu) (\partial_\mu B^\alpha)
\partial_\nu f_\alpha - (\delta V / \delta B_\nu) B_\alpha
\partial_\nu f_\alpha \label{last} \;.
\end{eqnarray}
Each one of the three terms in (\ref{last}) is separately zero: for
translations this is simply due to $\partial f_\mu = 0$, while for
Lorentz transformations each term is a product of a symmetric
expression and the antisymmetric $\omega_{\mu \nu}$. What is left
is then the expression in (\ref{secondlast}) which reads
\begin{equation}\label{djb}
\partial_\mu J_f^\mu = B_\alpha (-{\bf L}_f P^\alpha) = \Psi [P^\alpha] (- \delta_f P^\alpha) \;.
\end{equation}
Let us make here several comments:

(I) In general the Poincar\'e symmetry is broken because we cannot
implement the constraint $\Psi [P^\alpha] = 0$ unless we want that
the theory becomes trivial, $B_\alpha = 0$. One may argue that it
never seems meaningful to require $\Psi [\rm \chi_k ] = 0$, but it
is not so and this is at the heart of what in \cite{Iorio:2001qy}
is called dynamical consistency. The (counter-)example one could
consider is that of the dummy fields in supersymmetric theories,
as we shall show in some details later.  There is still room for
dynamical consistency, though, for noninvariant theories as the
theory (\ref{B}). In this case the charges are in general not
conserved because $\Psi[ \chi_k ] = 0$ does not make sense, but
they still generate the $\Delta$s and one has to demand that
$\Delta \phi_j = \delta \phi_j$ while $\Delta \chi_k = 0$.

(II) The algebraic transformations of the non-dynamical field
appear on the right side of (\ref{djb}) regardless of whether this
field has been varied or not in the action to obtain the current.
They are produced by the combination of $(\Psi [\phi_j ] = 0)
\times \delta_f \phi_j $ (term $B_\alpha P^\nu \partial_\nu
f^\alpha$ here) and of $f^\mu \partial_\mu {\cal L}$ (term $-
B_\alpha f^\mu \partial_\mu P^\alpha$ here).

(III) It is possible in this case to have relic symmetries, i.e.
to set $\delta_f \chi_k$ to zero without making $J_f^\mu$
trivially vanishing. Thus there is a sub-set of the parameters
$f_\mu$ for which there is invariance, namely the solutions to
${\bf L}_f P^\alpha = 0$. For translations ${\bf L}_f P^\alpha =
a^\mu
\partial_\mu P^\alpha = 0$, i.e. the directional derivative along
$a^\mu$ of $P^\alpha$ must vanish. For nonconstant $P^\alpha$ only
those translations are symmetries, hence, in general not even
$T^{\mu \nu}$ is always conserved. To have at least general energy
and momentum conservation one chooses a constant $P^\alpha$ which
gives as conditions for relic Lorentz symmetry $\omega^\alpha_\mu
P^\mu = 0$. This gives $\vec{\zeta} \cdot \vec{P} = 0$ and
$\vec{\omega} \times \vec{P} = P_0 \vec{\zeta}$, where $\omega^{0
i} = \zeta^i$ and $\omega^{i j} = \epsilon^{i j k} \omega_k$, with
$\vec{\zeta}$ the rapidity vector and $\vec{\omega}$ identifying
the axis of rotation. For $P_0 = 0$ all boosts in the plane
orthogonal to $\vec{P}$ and all rotations around $\vec{P}$ are
solutions, thus the subgroup of $SO(3,1)$ they identify is
$SO(2,1)$.

(IV) $\delta_f P^\alpha$ enter the expression for the conservation
of the current with the minus sign. As it does not make sense to
set $\Psi[P^\alpha]$ to zero, the flux of the current is, in
general, nonzero and proportional to the variations of the
back-ground field seen from the point of view of the transforming
field, i.e. transforming with the opposite sign. If, for instance,
the dynamical field rotates of an angle $\vartheta$ the relative
angular momentum has a net flux given by ($\Psi[P^\alpha]$ times)
a rotation of the background field of an angle $- \vartheta$. This
mechanism gives a precise meaning to what in literature is
sometimes referred to as the ``decoupling'' between ``particle''
and ``observer'' transformations occurring in certain (Lorentz)
noninvariant models, such as the Standard Model Extension (SME)
\cite{Colladay:1998fq}: From the point of view of the Noether
currents there is no ambiguity and always the invariance is broken
with the exception of the relic symmetries. Hence {\it only one
kind} of transformations is generated by the Noether charges (the
particle transformations) while the other transformations manifest
themselves as the terms breaking the invariance in the way
described above.

Let us explain now in more details why for dummy fields in
supersymmetric theories the constraint $\Phi [\chi_k] =0$ makes
sense. Although this is a general result, let us consider the
simple case of the massive free Wess-Zumino theory. The lagrangian
is
\begin{equation}\label{wz}
{\cal L}_{WZ} = - \partial_\mu \varphi \partial^\mu
\varphi^\dagger + D D^\dagger + [(-\frac{i}{2} \psi \not\!\partial
{\bar\psi} + m \varphi D - \frac{m}{2} \psi^2) + ({\rm h.c.})] \;,
\end{equation}
and the algebraic supersymmetry transformations that leave it
invariant are
\begin{eqnarray}
\delta \varphi = \sqrt2 \epsilon\psi & \delta \varphi^\dagger =
\sqrt2 \bar\epsilon \bar\psi \\
\delta \psi_\alpha  = i \sqrt2 (\sigma^\mu \bar\epsilon)_\alpha
\partial_\mu \varphi + \sqrt2 \epsilon_\alpha D & \quad \delta \bar{\psi}^{\dot{\alpha}} = i
\sqrt2 (\bar{\sigma}^\mu \epsilon)^{\dot{\alpha}} \partial_\mu
\phi^\dagger
+ \sqrt2 \bar\epsilon^{\dot{\alpha}} D^\dagger \\
\delta D = i \sqrt2 {\bar\epsilon}\not\!{\bar \partial} \psi &
\delta D^\dagger = i \sqrt2 \epsilon \not\!\partial {\bar\psi}
\end{eqnarray}
where $\varphi$ is the dynamical complex scalar field, $\psi$ is
its (Weyl) partner and $D$ is the nondynamical complex scalar
field. Here the implementation of the constraint $\Psi [D] =
D^\dag + m \varphi = 0$ (and its h.c.) gives a perfectly
meaningful theory, namely the free, massive Wess-Zumino lagrangian
\begin{equation}\label{wzonshell}
{\cal L}_{WZ} = - \partial_\mu \varphi \partial^\mu
\varphi^\dagger + m^2 \varphi \varphi^\dag -\frac{i}{2} ( \psi
\not\!\partial {\bar\psi}  - \bar{\psi} \not\!\bar{\partial} \psi
) - \frac{m}{2} ( \psi^2 + \bar{\psi}^2 ) \;.
\end{equation}
Furthermore, the supercurrent, $J^\mu_{\rm susy} = \sqrt{2}
(\bar{\psi} \bar{\sigma}^\mu \sigma^\nu \bar{\epsilon}
\partial_\nu \varphi - i m \epsilon \sigma^\mu \bar{\psi} \varphi^\dag + h.c.)$,
is conserved and the relative charge $Q_{\rm susy} = \int d^3 x
J^0_{\rm susy}$ generates on-shell also the transformations of
$D$, even though there is no associate momentum $\Pi_D$ to $D$.
This is easily seen by considering that $\Psi [D^\dag] = 0$ means
$D = - m \varphi^\dag$, while $\Psi [\bar{\psi}^{\dot\alpha}] = 0$
means $i (\not\!\!{\bar\partial} \psi)_{\dot\alpha} =  - m
{\bar\psi}_{\dot\alpha}$, thus acting on-shell with $Q_{\rm susy}$
on $D$ gives $\Delta_{\rm susy} D = \{ D , Q_{\rm susy} \} =  - m
\sqrt{2} \bar\epsilon \bar\psi \{ \varphi^\dag ,
\Pi_{\varphi^\dag} \} = - m \sqrt{2} \bar\epsilon \bar\psi$, with
$\Pi_{\varphi^\dag} = \partial_0 \varphi$. Using the on-shell
expression for $\bar\psi_{\dot\alpha}$, gives $\Delta_{\rm susy} D
= i \sqrt{2} {\bar\epsilon}\not\!{\bar
\partial} \psi$, which coincides with the algebraic
transformation. This is an illuminating instance of dynamical
consistency: supposing $\Psi[ \phi_j ] = 0$ is always implemented,
$\Psi[ \chi_k ] = 0$  on one side gives conservation, on the other
side gives an expression for the nondynamical field in terms of
dynamical ones $\chi_k (\phi_j)$ that, when acted upon with the
charge, gives back precisely the algebraic transformation: $\Delta
= \delta$.

With the help of the previous considerations, to treat the case in point of the NCGT (\ref{firstorder})
is now fairly easy. The current has the form $J^\mu_f = \Pi^{\mu \nu} \delta_f A_\nu - \hat{\cal L} f^\mu$ and
the divergence can be written as follows
\begin{eqnarray}
\partial_\mu J^\mu_f & = & \Pi^{\mu \nu} F_{\alpha \nu} \partial_\mu f^\alpha =
\left( \frac{\delta \hat{\cal L}}{\delta F_{\mu \nu}} F_{\alpha \nu} \right) 2 \partial_\mu f^\alpha \\
& = & \left( \theta^{\mu \beta} \frac{\delta \hat{\cal L}}{\delta \theta^{\alpha \beta}} \right) 2 \partial_\mu f^\alpha
= \frac{\delta \hat{\cal L}}{\delta \theta^{\alpha \beta}} \left( \theta^{\mu \beta}  2 \partial_\mu f^\alpha \right) \label{delta-1}\\
& = & \Psi[\theta^{\alpha \beta}] ( - {\bf L}_f \theta^{\alpha \beta} ) =
\Psi[\theta^{\alpha \beta}] ( - \delta_f \theta^{\alpha \beta} ) \label{deltatheta} \;.
\end{eqnarray}
Let us prove it. It was shown in \cite{Berrino:2002ss} that, after
partial integration, no derivatives of $F_{\mu \nu}$ appear in the
expansion hence one can write symbolically $\hat{{\cal L}} \sim
\sum_n \theta^n F^{n+2}$, i.e. the lagrangian is a homogeneous
polynomial in $\theta$ and $F$. Furthermore, only two things can
happen: either one given $\theta$ is coupled to one $F$ (i) or to
two $F$s (ii). Notice now that deriving $\hat{{\cal L}}$ w.r.t.
$F$ and then multiplying by $F$ produces precisely the same result
as multiplying by $\theta$ and then deriving w.r.t. $\theta$ (in
reverse order) because: in case (i) the $\theta^{\mu \nu}$ singled
out from the derivation $\delta / \delta F_{\mu \nu}$ contracts
(with the $\mu$ of $\partial_\mu f^\alpha$ and) with the $\nu$ of
the outcome of the derivation with $\delta / \delta \theta^{\alpha
\nu}$, i.e. $F_{\alpha \nu}$ times the same terms multiplying
$\theta^{\mu \nu}$; in case (ii) when the free index of $\theta$
left out of the derivation is $\nu$ then the contribution is zero
for a mechanism of cancelation we shall soon describe, while when
the free index is $\mu$ (say $\theta^{\mu \beta}$)  then the $\nu$
that contracts with the $F_{\alpha \nu}$ must be on one $F$, and
together with the $F_{\alpha \nu}$ gives what it would be obtained
by deriving with $\delta / \delta \theta^{\alpha \beta}$. There is
still to address the apparent mismatch between the number of terms
produced by deriving $\sum_n \theta^n F^{n+2}$ w.r.t. $F$ and the
number of terms obtained by deriving it w.r.t. $\theta$. They are
in fact the same. For translations everything vanishes. For
Lorentz transformations $\partial_\mu f^\alpha =
\omega^\alpha_\mu$. Let us consider this case. The extra two terms
one obtains by deriving w.r.t. $F$ vanish because: either one gets
$\sim 2 (\theta^n F^n) (F^{\mu \nu} F_{\alpha \nu})
\omega^\alpha_\mu$ (case (i) above) or one gets $\sim [(\theta^n
F^{n+2})^\mu_{\,\, \alpha} + (\theta^n F^{n+2})^{\,\,\,
\mu}_\alpha ] \omega^\alpha_\mu$ (case (ii) above), i.e. always a
symmetric expression times $\omega^\alpha_\mu$. The latter
cancelation is also responsible for the matching $(\delta / \delta
F) F \sim \theta {\delta / \delta \theta}$ in case (ii) above.
Finally, notice that for constant $\theta^{\alpha \beta}$:
$A_{\alpha \beta} {\bf L}_f \theta^{\alpha \beta} = A_{\alpha
\beta} \theta^{\mu \alpha} 2 \partial_\mu f^\beta $ for any
antisymmetric $A_{\alpha \beta}$. Collecting all this information
gives the result (\ref{deltatheta}).

The discussion on invariance and dynamical consistency goes along
the lines of the previous discussion. The theory is in general
{\it not invariant} under Poincar\'e transformations because it
does not make sense to set $\Psi[\theta^{\alpha \beta}]$ to zero:
on the one hand this constraint would make the theory trivial
(e.g. at first order we would get $F_{\mu \nu} = 0$), on the other
hand it does not allow to express $\theta (F)$ in a meaningful
way. That is why the only way left for dynamical consistency is
$\Delta_f \theta_{\mu \nu} = 0$ and $\Delta_f A_\mu = \delta_f
A_\mu$, as proved already in \cite{Iorio:2001qy}. There is room
for relic symmetries found by solving ${\bf L}_f \theta^{\mu \nu}
= 0$, which is satisfied for all translations. For Lorentz
transformations we have to solve $\theta^{\mu \beta}
\omega^\alpha_{\;\; \mu} = 0$. For $\beta = 0$ we get
$\vec{\tilde{\theta}} \times \vec{\omega} = 0$, while for $\beta =
j$ and $\alpha = 0$ we get $\vec{\theta} \times \vec{\zeta} = 0$,
where $\vec{\omega}$ and $\vec{\zeta}$ have been already defined,
$\vec{\tilde{\theta}} = (\theta^{0 1}, \theta^{0 2}, \theta^{0
3})$ and $\vec{\theta} = (\theta^1, \theta^2, \theta^3)$, with
$\theta^{i j} = \epsilon^{i j k} \theta_k$. Furthermore, by taking
the equation for $\beta = j$ and $\alpha = k$, $\tilde{\theta}^j
\zeta_k = - \delta^j_k \vec{\theta} \cdot \vec{\omega} + \theta_k
\omega^j$, and contracting it with $\omega_k$ we get $\vec{\zeta}
\cdot \vec{\omega} = 0$, while contracting it with $\theta_j$ we
get $\vec{\tilde{\theta}} \cdot \vec{\theta} = 0$. Thus, rotations
around $\vec{\tilde{\theta}}$ and boosts along $\vec{\theta}$ are
still symmetries, provided $\vec{\tilde{\theta}} \cdot
\vec{\theta} = 0$. The group is $SO(2) \times SO(1,1)$, modulo
some discrete symmetries, and it has been considered in great
detail in \cite{Alvarez-Gaume:2003mb}.

Note that the expression (\ref{delta-1}) coincides, at first order
in $\theta$, with what is obtained in \cite{Banerjee:2004ev} (use
$\hat{\cal L}|_{O(\theta)}$ in (\ref{firstorder}), compute the
derivatives and rearrange the terms as discussed) and interpreted
there as ``the criterion for Lorentz invariance in the case
$\theta_{\mu \nu}$ transforms like a tensor'', which is evidently
an incorrect interpretation (cf. Eqs. (81) and (82) and also Eq.
(87) and the following discussion).

We conclude that for (abelian) NCGTs of the kind in (\ref{firstorder}) only relic
symmetries are present and they are: translations in any direction and
the subgroup of the Lorentz group compatible with $\theta_{\mu \nu}$, i.e.
whose parameters satisfy ${\bf L}_f \theta_{\mu \nu} = 0$ ($SO(2) \times SO(1,1)$ for
the case $\vec{\tilde{\theta}} \cdot \vec{\theta} = 0$). There is no
choice on whether to transform or not transform $\theta_{\mu \nu}$: for dynamical
consistency $\Delta_f \theta_{\mu \nu} = 0$ for all $f^\mu$, while the
$\delta_f \theta_{\mu \nu}$ are precisely the terms breaking the invariance
in general (as $\Psi [\theta_{\mu \nu} ] = 0 $ cannot be implemented) and they are
produced, with the minus sign, in $\partial_\mu J^\mu$ regardless of whether
$\theta_{\mu \nu}$ has been varied in the action due to the particular
coupling of these fields with the dynamical field $F_{\mu \nu}$. This means that
the system undergoing a Lorentz transformation with parameter $\omega^\alpha_\mu$, not
belonging to the relic symmetries, sees the $\theta_{\mu \nu}$ transforming
with the parameter $- \omega^\alpha_\mu$ as the cause of nonconservation of the $M^{\mu \nu \lambda}$.
This also solves the ambiguity of the ``particle'' and ``observer'' transformations in this
context, as the Noether (in general not conserved) charges only can generate one kind
of transformations (particle) while the other kind (observer) are seen to appear in the
way illustrated above. Thus, there is only one criterion for Lorentz invariance of
NCGTs and it is the usual one $\partial_\mu M^{\mu \nu \lambda} = 0$.

We hope that this Comment will ultimately clarify that for NCGTs of the kind discussed here {\it standard}
Lorentz symmetry is not present and that statements like ``NGCT has Lorentz
invariance only when $\theta_{\mu \nu}$ transforms like a tensor'' are simply wrong.
We did not consider here noncommutative {\it modifications} of the Lorentz algebra - as, e.g.,
that proposed in \cite{Chaichian:2004za} (the twisted coproduct
approach) or in \cite{Calmet:2004ii} (the $\theta$-{\it deformed}
transformations approach) - where the transformations themselves
are modified hence a different meaning must be ascribed to Lorentz
invariance. In \cite{Banerjee:2004ev} and in \cite{Iorio:2001qy} the same
approach is used of standard Lorentz transformations hence there is no room for
opposite conclusions on Lorentz invariance.

One further development of this analysis is to prove under which conditions the following conjecture
is true: When in the action the nondynamical $\chi^1_{k_1}, ..., \chi^n_{k_n}$ are coupled to the
dynamical $\phi^1_{j_1}, ..., \phi^m_{j_m}$ (and/or their derivatives) to obtain what
\textit{would be a scalar} if both sets of fields are transformed \textit{algebraically} then
the result
\begin{equation}\label{main}
\partial_\mu J_f^\mu = \sum_{i=1}^n \Psi [\chi^i_{k_i}] (- \delta_f \chi^i_{k_i}) \;,
\end{equation}
holds. Here $\sum_{i=1}^m \Psi [\phi^i_{j_i}] = 0$ has been used
and let us stress again that the expression of the current on the
left side is independent on whether the non-dynamical fields in
the action have been varied or not. We expect that this is the
case of the SME \cite{Colladay:1998fq}. Finally, it would be
interesting to study within this approach the supercurrents of the
Lorentz-violating Wess-Zumino model proposed in the SME context in
Ref. \cite{Berger:2001rm}.

{\bf Acknowledgments}

We thank Zuzana Vydorova for valuable help with some computations.

 \end{document}